\documentclass[epj,english,final,amsmath,amssymb]{svjour}
\usepackage{graphicx}
\input{epsf}
\begin{document}
\title{Photogalvanic current in artificial asymmetric nanostructures}

\author{A. D. Chepelianskii\inst{1}, M. V. Entin\inst{2}, 
L. I. Magarill\inst{2}  and D. L. Shepelyansky\inst{3,}\inst{4}}

\institute {Ecole Normale Sup\'erieure, 45, rue d'Ulm, 
75231 Paris Cedex 05, France
\and
Institute of Semiconductor Physics, 
Siberian Division of Russian Academy of Sciences,
Novosibirsk, 630090, Russia
\and
Laboratoire de Physique Th\'eorique,
UMR 5152 du CNRS, Univ. P. Sabatier, 31062 Toulouse Cedex 4, France
\and
http://www.quantware.ups-tlse.fr/dima/}

\titlerunning{Photogalvanic current in artificial asymmetric nanostructures}
\authorrunning{A.D.Chepelianskii {\it et al.}}

\date{Received:}
\abstract{We develop a theoretic description of the photogalvanic
current induced by a high frequency radiation
in asymmetric nanostructures 
and show that it describes well the results of numerical simulations.
Our studies allow to understand the origin of the electronic
ratchet transport in such systems and show that they can be used
for creation of new types of detectors operating
at room temperature in a terahertz radiation range.
}

\PACS{
{72.40.+w}{Photoconduction and photovoltaic effects}
\and
{73.63.-b}{Electronic transport in nanoscale materials and structures} 
\and
{05.45.Ac}{Low-dimensional chaos}
}
\date{January 7, 2007}

\maketitle


\section{Introduction}

Since the experiments of Glass {\it et al.} 
performed in 1974 \cite{glass1974}
and the theoretical explanations 
developed in 1976 \cite{entin1976,belinicher1976}
it is known that an asymmetry of crystal at a microscopic
scale can lead to emergence of macroscopic stationary
directed current  when the crystal is irradiated by an external
light source. The appearance of directed current
induced by zero-mean force of radiation
in absence of any external static voltage 
has been named the photogalvanic effect.
This unusual phenomenon based on interplay of
space asymmetry, relaxation and external driving
had been mainly discussed for interaction
of light with crystals \cite{entin1978,belinicher1980}.
However, this effect has a rather generic physical origin
and the interest to it has been significantly renewed 
recently   when it became clear that it may play
an important role for transport in bio-systems
where it is difficult to create static forces
in space and where directed transport may be 
more easily generated by some oscillating
parameters in presence of asymmetry at a molecular level
(see {\it e.g.} a review \cite{prost}).
In this community the phenomenon
became known as ratchet, following an example of pawl 
and ratchet described by Feynman and showing that
a directed transport in asymmetric systems
at a thermal equilibrium
is forbidden by the second law of thermodynamics
\cite{feynman}. 

The appearance of the photogalvanic or ratchet 
effect in various systems is described in
the reviews \cite{hanggi,reimann}.
The effect has been observed with vortices
in Josephson junction arrays \cite{mooij,nori,ustinov},
cold atoms \cite{grynberg}, macroporous silicon membranes \cite{muller},
microfluidic channels \cite{ajdari} and other systems.
The great variety of systems clearly confirms a generic nature of
the phenomenon.
 
In parallel to these ratchet studies, a technological
progress made possible to produce artificial
superlattices in semiconductors with 
two-dimensional electron gas $\;\;$ (2DES).
The experimental studies of superlattices
of antidots in a form of disks
demonstrated an important contribution
of periodic orbits in the transport
properties of 2DES \cite{weiss,kvon}.
It is important to note that the interest
to particle dynamics on a lattice of rigid
disks goes back to the days of Galton who in 
far 1889 showed the appearance of statistical
laws in such systems \cite{galton}.
A rigorous mathematical description
of chaos on the Galton board
that leads to statistical laws
has been given by Sinai \cite{sinai}.
The links between the chaotic dynamics,
periodic orbits and experimental results
for 2DES transport properties
in antidot superlattices have been
established in theoretical studies \cite{geisel}. 
The interest to effects of microwave radiation
on  2DES transport appeared at a relatively
early stage, however, 
during a relatively long time only
the case of antidots with a disk shape
has been considered 
(see {\it e.g.} \cite{kvon1,weiss1}).
Due to symmetry reasons the photogalvanic 
effect is absent in such a case.

The experimental studies of 2DES transport
in asymmetric structures in presence of
{\it ac-}driving have $\:\:$ been started in
\cite{lorke,linke,song}. They demonstrated 
the principal existence of photogalvanic transport
but no detailed analysis had been done for
2D structures. The quasi-1D case \cite{linke} has
been analyzed in more detail 
but the ratchet transport in this
case is rather slow due to 
very slow {\it ac-}driving. 
Thus, the ratchet transport in this case was linked
to quantum tunneling effects 
(see \cite{linke} and  discussions
in \cite{reimann}).
On the contrary the experiments in 2D structures
\cite{lorke,song} have been performed
at rather high frequencies (50GHz and more)
showing that high frequency control
of ratchet transport is possible in principle.
Moreover, the experiment \cite{song}
demonstrated that the photogalvanic current exists
at room temperature.
Unfortunately, there was no further
development of this interesting research line.
Probably, the absence of theoretical understanding
of the phenomenon is partially responsible for this.

The theoretical studies of ratchet transport
in asymmetric 2DES structures in the form 
of semidisks Galton board
have been started in \cite{alik2005}
and further developed in
\cite{cristadoro,alik2006}. They used
extensive numerical simulations of dynamical equations
in combination with simple analytical estimates.
A rigorous analytical approach,
based on the kinetic equation,
has been developed in \cite{entin2006}
that allowed to solve exactly a case
at a low density of 
asymmetric scatterers of a specific form.
Here we combine all these methods
that allows us to obtain a global
theoretical description of the photogalvanic
effect in asymmetric nanostrutures.
The development of such global theory
allows to make clear predictions
for conditions under which
the photogalvanic effect can be used
for construction of room temperature
detectors of high frequency radiation
sensitive to polarization.

The paper has the following structure:
Section II gives a description of
various  models of asymmetric nanostructures
and represents simple analytical computations
compared with numerical simulations;
Section III describes analytical results
based on the kinetic equation
and also considers a general question of ratchets
in dynamical systems with or without time reversibility;
the effects of magnetic field
on the photogalvanic current are
considered in Section IV; 
discussion of the results
and possible application of the effect
are given in Section V.

\section{Model description, analytical and numerical results}

\begin{figure}
\vglue 0.4cm
   \centering
   \includegraphics[height=0.25\textwidth,angle=0]{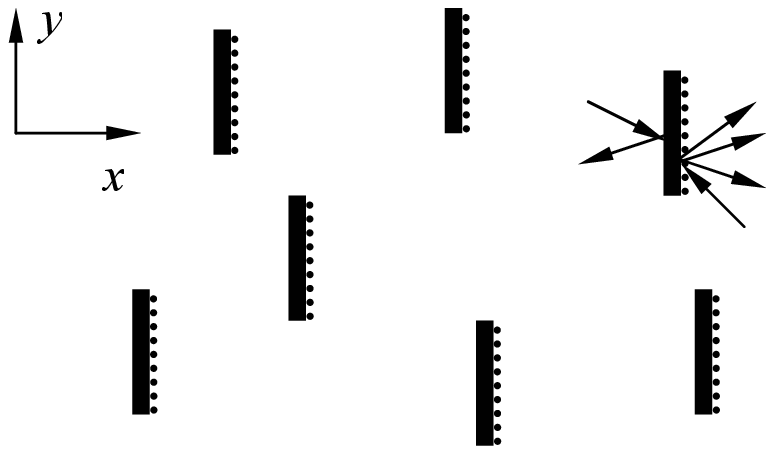}
\vglue 0.5cm
\caption{The model of oriented scatterers in a form of vertical
cuts of length $D$ with a concentration $n_c$.
The scattering on cuts is elastic from the left side
and diffusive from the right side.
The average scattering time on cuts is $\tau_c=1/(n_c D V_F)$,
where $V_F$ is the Fermi velocity of 2DES.
}
\label{fig1}
\end{figure}

\begin{figure}
\vglue 0.5cm
\vglue 0.4cm
   \centering
   \includegraphics[height=0.33\textwidth,angle=0]{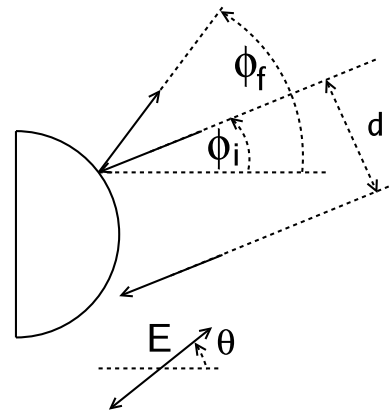}
\caption{Geometry of the semidisk scatterer 
in $(x,y)-$plane and definition of the 
angle notations. A trajectory colliding with the semidisk at 
an angle $\phi_i$ is scattered at an angle $\phi_f$, which depends 
on the impact parameter $d$; the electric field 
$E$ is linear-polarized under angle $\theta$
to $x-$axis.
}
\label{fig2}
\end{figure}

In our studies we consider two main types of
antidots: oriented elastic semidisks and 
cuts (1D intervals of length D)
which produce specular reflection
from left side and diffusive scattering
reflection from right side
(see Fig.~\ref{fig1}). The cuts model
has been introduced in \cite{entin2006}
to mimic effect of scattering on a semidisk
(right diffusive side of cut approximately represents
circular part of semidisk). The kinetic equation can be solved
exactly in this case. For the semidisks Galton board
we assume that  the semidisks form a triangular
lattice (see Fig.1 in \cite{alik2006})
with $R$ being a distance between disk centers 
and $r_d$ being the disk radius. 
Orientation of a semidisk on $(x,y)-$plane 
and angles of elastic scattering are shown in Fig.~\ref{fig2}.
For the case of cuts,
as in \cite{entin2006},
it is assumed that the cuts are irregularly distributed in space
with a concentration of cuts in a unit area being $n_c$
(all cuts are vertical as in Fig.~\ref{fig1}).
The kinetic theory \cite{entin2006} also works in a case of
regular lattice of cuts if their density is low.

In the limit of low density of scatterers (cuts or disks)
the scattering time $\tau_c$ is 
\begin{equation}
\label{eq1}
      \begin{array}{ll}
\tau_c=1/(n_c D v) \;\; ({\bf cuts}), \\
\tau_c=\sqrt{3} R^2/(8 r_d v) \;\; ({\bf semidisks}) ,
       \end{array}
\end{equation}
where $v$  is the particle velocity
(for the Fermi gas $v=V_F$ where $V_F$ is the Fermi velocity of 2DES).
In addition to scattering on antidots we assume that
there is also scattering on impurities
which is characterized by the scattering time $\tau_i$
(see \cite{alik2006,entin2006}). Between the collisions
with antidots and impurities the electron motion
is affected only by an  electric microwave field
$\mathbf{E} \cos \omega t$ linear-polarized under angle
$\theta$ to $x-$axis (see Fig.~\ref{fig2}).
The force acting on electron is
$\mathbf{F} = e \mathbf{E}$ and the electron velocity
at time $t+\tau_s$ is
\begin{equation}
\label{eq2}
\mathbf{v}(t+\tau_s) = \mathbf{v}(t) + 
\frac{2 e\mathbf{E}}{m \omega} 
\cos(\omega t + \omega\tau_s / 2) \sin(\omega \tau_s / 2) ,
\end{equation}
where $e, m$ are electron charge and mass.

To find the velocity of stationary flow 
induced by microwave oscillations we assume
that without microwave the equilibrium distribution
of electron velocities is given by the Maxwell
or the Fermi-Dirac distributions  at temperature $T$.
Also it is assumed that the microwave field is relatively weak
and it only slightly perturbs the equilibrium distribution.
Let us start for simplicity from the case
of the  Maxwell distribution $f_M(\mathbf{v})=\exp(- m \mathbf{v}^2/2T)/Z$
(here $Z$ is the normalization constant).
Then, according to (\ref{eq2}), the perturbed distribution is 
\begin{eqnarray}
\nonumber
&&f(\mathbf{v},t,\tau_s) = f_M(\mathbf{v}) \times \\
\label{eq3}
&&\exp[{\frac{2 e \mathbf{v} \mathbf{E} 
\cos(\omega t + \omega \tau_s / 2) \sin(\omega \tau_s / 2)}{T \omega}}] \; ,
\end{eqnarray}
where we omit the velocity independent terms.

To compute the ratchet velocity we first consider the cuts model.
We also make certain simplifications which allow to understand
the physical origin of the ratchet transport
keeping a more rigorous treatment for the next Section.
Thus, we compute the averaged perturbed distribution function.
It is obtained by averaging $f(\mathbf{v},t,\tau_s)$ 
over all initial times $t$ and 
the free flight time $\tau_s$.
\begin{equation} 
\label{eq4}
f_c(\mathbf{v}) = <\int_0^{\infty} p_c(\tau) f(\mathbf{v}, t, \tau_s) d \tau_s >_t 
\end{equation}

Here $p_c(\tau_s)$ is the probability that the scattering occurs 
after a time $\tau_s$.
Its exact expression depends on the geometry of the sample and  temperature, 
but for simplicity we may assume that 
$p_c(\tau_s) = \exp(-\tau_s/\tau_c) / \tau_c$ where 
$\tau_c$ is the mean scattering time.
Expanding the exponent up to a second order in  
$\mathbf{E}$ and averaging over $t, \tau_s$ gives: 
\begin{eqnarray}
\nonumber 
&&f_c(\mathbf{v}) =  f_M(\mathbf{v}) \times \\
\label{eq5} 
&&\left(1 + \frac{(\tau_c  \mathbf{v} e \mathbf{E} )^2 }{T^2} 
\frac{1}{ 1 + (\tau_c \omega)^2 } \right)
\end{eqnarray}

With this distribution we can compute the ratchet
flow velocity $v_{f,x}$ (in x direction) 
considering only one scattering on a cut. Indeed, 
$v_{f,x}$ is the sum of two terms originated from 
scatterings on the two sides of a cut $v_{f,x} = v_{+,x} - v_{-,x}$.   
The contribution from the right side is $v_{+,x} = <|\mathbf{v}|> / \pi$ 
since the outgoing direction of the particle is random and 
the average depends only on the magnitude of the velocity 
before scattering $|\mathbf{v}|$. 
On the contrary the scattering on the left side just changes the sign 
of the velocity in the x direction: $v_x$, as a consequence 
$v_{-,x} = < v_x \eta(v_x) >$ where $\eta(v_x)$ is the step theta-function. 
The $1/\pi$ factor in the expression of $v_{+,x}$ originates from the averaging 
over the random scattering angle, it is not present for $v_{-,x}$ since 
the scattering on the left side of the cut is elastic and no additional 
averaging is needed.
If the distribution is isotropic these two averages are equal, 
keeping this fact in mind 
we may compute the contribution to $v_{+,x}$, $v_{-,x}$ 
only from the anisotropic (non equilibrium)  
part of $f_c(\mathbf{v})$. This gives 
\begin{eqnarray}
\nonumber
&&v_{f,x} = v_{+,x} - v_{-,x} 
    = \frac{3 \tau_c^2 e^2 E^2}{ 2 \sqrt{2 \pi m^3 T} } \frac{1}{ 1 + (\tau_c \omega)^2 } \times \\
\nonumber
&& \left( \frac{1}{\pi} \int_0^{2 \pi} \cos^2( \phi - \theta) d \phi - 
       \int_{-\pi /2 }^{\pi / 2} \cos^2(\phi - \theta) \cos(\phi) d \phi \right) \\
&& = -\frac{\tau_c^2 e^2 E^2}{ 2 \sqrt{2 \pi m^3 T} } 
       \frac{1}{ 1 + (\tau_c \omega)^2 } \cos(2 \theta) \; ,
\label{eq6}
\end{eqnarray}
where $\phi$ is the polar angle of the velocity $\mathbf{v}$, 
$\theta$ is the polarization angle and the coefficient comes from the average value of
$<v^4>$ computed over the equilibrium distribution $f_M$. 

Similar computations give the  ratchet velocity $v_{f,y}$ in $y-$direction.
Here the contribution from the right cut side vanishes $v_{+,y} = 0$ since after a random 
scattering the directions with $\phi$ and $-\phi$ have equal probability.
The contribution from the left side is given by $v_{-,y} = <v_y \eta(v_x)>$:
\begin{eqnarray}
\nonumber
&&v_{f,y} = v_{-,y} = \frac{3 \tau_c^2 e^2 E^2}{ 2 \sqrt{2 \pi m^3 T} } 
   \frac{1}{ 1 + (\tau_c \omega)^2 } \times \\
\nonumber
&&\left(\int_{-\pi /2 }^{\pi / 2} \cos^2(\phi - \theta) \sin(\phi) d \phi \right) \\
&&= \frac{\tau_c^2 e^2 E^2}{ \sqrt{2 \pi m^3 T} } \frac{1}{ 1 + (\tau_c \omega)^2 } \sin(2 \theta)
\label{eq7}
\end{eqnarray}

As a result Eqs.~(\ref{eq6},\ref{eq7}) give the angle $\psi$ of the direction of ratchet flow 
($\tan \psi = v_{f,y}/v_{f,x}$) as a function of polarization angle $\theta$:
\begin{equation}
\tan( \psi ) = -2 \tan(2 \theta) 
\label{eq8}
\end{equation}
It is important to note that in the computations above we assumed that 
a scattering event definitely occurs after time $\tau_c$. This model is rather
convenient for numerical simulations that allows to make 
a comparison with the above theoretical estimates.
Such a model corresponds to scattering events randomly placed in time.
For a static random distribution of cuts in space
one should compute the scattering 
probability with a transport cross section that gives a numerical factor 3
instead of 2 in Eq.~(\ref{eq8}) (see next Section).

Also above it is assumed that there are no impurities.
In their presence the result is proportional to the probability of
scattering on antidots which is equal to the ratio $\tau_i/(\tau_i+\tau_c) = \tau/\tau_c$
due to time ergodicity. Here and below
$\tau=\tau_c \tau_i/(\tau_c+\tau_i)$ is the relaxation time
scale determined by the geometrical mean of
$\tau_i$ and $\tau_c$. Hence, in presence of impurities
the ratchet velocity is given by 
\begin{equation}
 \mathbf{v_f} = \frac{\tau^3 e^2 E^2}{ 2 \tau_c \sqrt{2 \pi m^3 T} } 
 \frac{1}{ 1 + (\omega \tau)^2 } 
 \left( \begin{array}{c} 
- \cos(2 \theta) \\ 
2 \sin(2 \theta) 
\end{array} \right)
\label{eq9}
\end{equation}
where up and down terms correspond to $x$ and $y$ components of the ratchet velocity.

The method described above can be also applied to another form of antidots,
{\it e.g.} for semidisks. The main difference from the cut model is that the direction 
after collision $\phi_f$ is no longer completely determined by the impact angle 
$\phi_i$ nor completely random (see Fig.~\ref{fig2}). 
In fact, it is given by a conditional probability 
$g(\phi_f | \phi_i)$ that depends on the geometry of the scatterer. 
Due to the scatterer asymmetry, in general,  this distribution is not 
invariant under $y$-axis mirror symmetry. To obtain the 
expression of $g(\phi_f | \phi_i)$ for a semidisk scatterer
it is convenient to represent $g$ via two parts: 
\begin{equation}
g(\phi_f|\phi_i) = \left\{
\begin{array}{lr}
g_+(\phi_f|\phi_i), \;\; \phi_i \in (-\pi/ 2, \pi / 2)\\ 
g_-(\phi_f|\phi_i), \;\; \phi_i \in (\pi/ 2, 3\pi / 2)
\end{array}
\right.
\label{eq10}
\end{equation}

Let us first compute the contribution to $g$ from collisions with negative 
$x$ impact velocity $g_+$ that
correspond to impact angles $\phi_i \in (-\pi/2, \pi/2)$.
The scattering is elastic on all sides of the semidisk. 
Integrating over the impact parameter $d$ (see Fig.~\ref{fig2}) 
gives the conditional probability $g_{+}(\phi_f | \phi_i)$
of scattering in the direction $\phi_f$ assuming a collision with 
an impact angle $\phi_i$. 
\begin{equation}
g_{+}(\phi_f | \phi_i) = \frac{ \cos \frac{\phi_f - \phi_i}{2} } { 2 ( 1 + \cos \phi_i ) } 
                        \chi_{[\phi_i - \pi, \pi - \phi_i]}(\phi_f)
\label{eq11}
\end{equation}
Here $\chi_{[\phi_i - \pi, \pi - \phi_i]}$ is the characteristic theta-function 
of the interval $[\phi_i - \pi, \pi - \phi_i]$. 
After scattering with an impact angle $\phi_i$ some values of the outgoing angle $\phi_f$ are 
forbidden since the trajectory can not cross the bulk of the scatterer, for example 
$\phi_i + \pi$ is always forbidden, the $\chi$ function incorporates 
this restriction into $g_{+}(\phi_f | \phi_i)$.

The contribution from the trajectories with positive $x$ impact velocities 
is given by the distribution $g_{-}(\phi_f | \phi_i)$ for impact angles in the interval 
$\phi_i \in (\pi/2, 3 \pi/2)$. The resulting conditional probability $g_{-}(\phi_f | \phi_i)$ 
splits in two terms: the first comes from the collision with the straight edge 
of the semidisk and is expressed as a delta function, the second is related 
to the collisions with the curved edge and is similar to the expression given 
in Eq.(\ref{eq11}). Hence,
\begin{eqnarray}
\nonumber
&&g_{-}(\phi_f | \phi_i) = \frac{2 |\cos \phi_i|}{ 1 + |\cos\phi_i| } \delta(\phi_f + \phi_i -2 \pi) \\
&&     + \frac{ |\cos \frac{\phi_f - \phi_i}{2}| } { 2 ( 1 + |\cos \phi_i| ) }   
           \chi_{[ -|\pi - \phi_i|, |\pi - \phi_i|]}(\phi_f)
\label{eq12}
\end{eqnarray}
One can check that the distributions obtained in Eqs.(\ref{eq11}, \ref{eq12}) are normalized to 1. 
This normalization corresponds to the conservation of the number of particles after collision:
$\int_{0}^{2 \pi} g_{+}(\phi_f | \phi_i) d\phi_f = \int_{0}^{2 \pi} g_{-}(\phi_f | \phi_i) d\phi_f = 1
\label{gnorm}$.

Another effect that was not taken into account in the cuts model
is that the scattering probability  depends on the impact angle $\phi_i$. 
The probability $g_c(\phi_i)$ that the collision 
occurs under the angle $\phi_i$ is proportional to the length of 
the segment obtained by projecting the semidisk on a parallel to 
the impact direction : $r_d (1 + |\cos(\phi_i)|)$, this leads 
to the probability distribution
\begin{equation}
g_c(\phi_i) = \frac{1+|\cos \phi_i|}{2(2 + \pi)}
\label{eq13}
\end{equation}
The distributions given by Eqs.~(\ref{eq11},\ref{eq12},\ref{eq13}) and the distribution function
$f_c(\mathbf{v})$ (\ref{eq5}) give the ratchet velocities via expressions:
\begin{eqnarray}
\nonumber
&&v_{f,x} = \int_{0}^{2 \pi} d\phi_f \int d^2 \mathbf{v}  |\mathbf{v}| \cos(\phi_f) g(\phi_f | \phi_i) g_c(\phi_i) f_c(\mathbf{v}) \\
\nonumber
&&v_{f,y} = \int_{0}^{2 \pi} d\phi_f \int d^2 \mathbf{v}  |\mathbf{v}| \sin(\phi_f) g(\phi_f | \phi_i) g_c(\phi_i) f_c(\mathbf{v}) \\
\label{eq14}
\end{eqnarray}

In the calculation of these integrals it turns out that the isotropic (equilibrium) term of 
$f_c(\mathbf{v})$ vanishes that corresponds to the absence of the effect at equilibrium.
In the contribution of the anisotropic term of Eq.(\ref{eq5}) only the angular integrals 
are different from Eq.~(\ref{eq6},\ref{eq7}) while the integral on $|\mathbf{v}|$ is identical 
and leads to the same dependence on system parameters. As a result we obtain 
\begin{eqnarray}
\nonumber
 \mathbf{v_f} = \frac{\pi \tau^3 e^2 E^2}{ 2 (2+\pi) \tau_c \sqrt{2 \pi m^3 T} } 
 \frac{1}{ 1 + (\omega \tau)^2 } 
 \left( \begin{array}{c} 
- \cos(2 \theta) \\ 
  \sin(2 \theta) 
\end{array} \right) \\
\label{eq15}
\end{eqnarray}

\begin{figure}
\vglue 0.5cm
\vglue 0.4cm
   \centering
   \includegraphics[height=0.33\textwidth,angle=0]{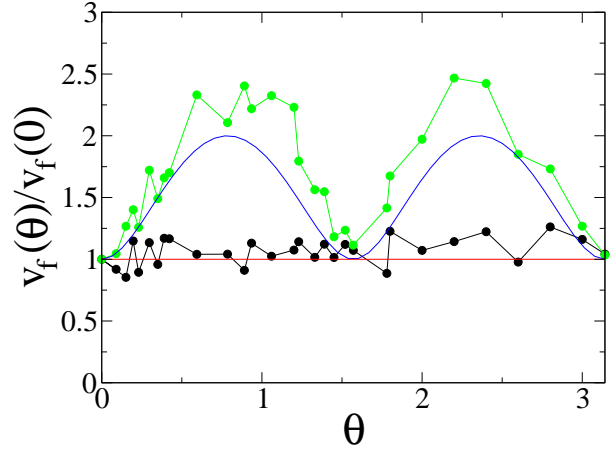}
\caption{Dependence of the rescaled absolute value of ratchet velocity 
$v_f(\theta)/v_f(\theta = 0)$ 
on the polarization angle $\theta$ for the cuts and semidisks models
(angle is given in radians).
For the the semidisk scatterers 
(black curve with circular dots) the lattice constant is set to $R/r_d = 4.5$,
the impurity scattering time is $\tau_i \approx 15 r_d/v_T$ and the Maxwell
equilibrium distribution is characterized by a temperature $T$. 
With these settings the numerically determined 
average time interval between two successive collisions 
with semidisks is $\tau_c \approx 0.77 \tau_i \approx 11 r_d/v_T$
(we note that 
the  theoretical scattering time is  $\tau_c \approx 4.4  r_d/v_T$
with the thermal velocity $v=v_T=\sqrt(2T/m)$ in Eq.~(\ref{eq1})).
In the simulations for the cuts model (green curve with circular dots) the 
temperature $T$ and the time scales $\tau_c$, $\tau_i$ are 
taken to be the same as their numerical values in the semidisks case.
In both cases the microwave field amplitude and frequency are
$eE r_d/T = 0.4$ and $\omega r_d/\sqrt{2mT} = 0.316$.
The smooth curves represent the theoretical predictions 
of Eq.~(\ref{eq9}) (blue curve for the cuts model) and Eq.~(\ref{eq15}) 
(red line for the semidisks model).
}
\label{fig3}
\end{figure}

\begin{figure}
\vglue 0.5cm
\vglue 0.4cm
   \centering
   \includegraphics[height=0.33\textwidth,angle=0]{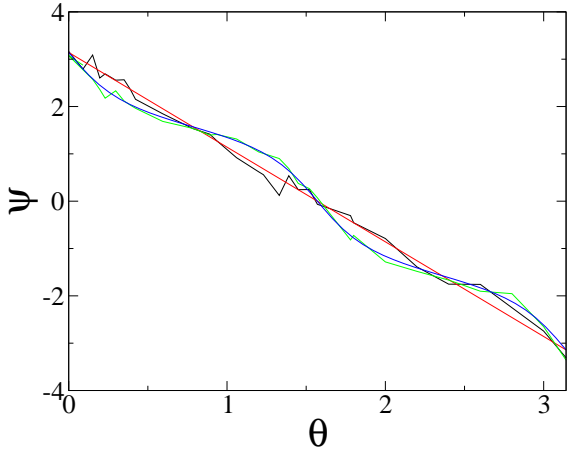}
\caption{Dependence of the angle $\psi$ of the ratchet flow direction
on the polarization angle $\theta$ for the parameters of Fig.~\ref{fig2}
with the same choice of colors: black curve for numerical data and red line
for theory ($\psi=\pi-2\theta$) 
in the semidisks model;
green curve for numerical data and blue curve for theory (see Eq.~(\ref{eq8}))
in the cuts model; angles are given in radians.
}
\label{fig4}
\end{figure}

To check the obtained theoretical expressions (\ref{eq9}) and (\ref{eq15})
we performed numerical simulations of the cuts and semidisks models.
The dynamical equations are solved numerically between collisions.
The  Maxwell equilibrium at temperature $T$ is generated with the help
of the Metropolis thermalization algorithm as it is described in \cite{alik2006}.
The computation time along one trajectory is about few
hundred thousands of microwave periods.
The angular dependence of the ratchet velocity is shown in Figs.~\ref{fig3},\ref{fig4}.
It is in a very good agreement with the obtained theoretical expressions
both for the cuts and semidisks models.
We will consider the dependence on the parameters $\tau_c, \tau_i$
in the next Section.

\section{The kinetic equation approach}

The approach described above can be also used for the Fermi-Dirac equilibrium
distribution. However, it if more convenient to use a more 
general approach based on the kinetic equation which reads:
\begin{equation}
\frac{\partial f}{\partial t} + e \mathbf{E} \cos {\omega t} 
\frac{\partial f}{\partial \mathbf{p}} 
 = - \frac{f - f_0}{\tau} \;  ,
\label{eq16} 
\end{equation}
where $f_0$ is an unperturbed equilibrium distribution, 
$\mathbf{p} = m\mathbf{v}$.
The solution can be presented in a form of expansion over 
powers of external weak field: $f = f_0 + f_1 + f_2 + ...$.
The first term is
\begin{equation}
f_1(\mathbf{p},t) = - \frac{\tau e \mathbf{E} \mathbf{v}}{ 2 (1 + i \omega \tau) } \frac{\partial f_0}{\partial \epsilon} \exp(i \omega t) + CC \; , 
\label{eq17}
\end{equation} 
where $CC$ is a complex conjugated part. The time averaged correction
of $f_2$ gives
\begin{eqnarray}
\nonumber
&& < f_2(\mathbf{p},t) >_t = - \tau e \mathbf{E} < \cos \omega t  \frac{\partial f_1}{\partial \mathbf{p}} > \\
&& \approx \frac{(\tau e \mathbf{E} \mathbf{v})^2 }{ 2 (1 + (\omega \tau)^2) } \frac{\partial^2 f_0}{\partial \epsilon^2} \; ,
\label{eq18}
\end{eqnarray}
where we use an approximation that  the isotropic term originating 
from the term $\partial f_1/\partial \mathbf{v}$ 
does not contribute to the ratchet velocity and therefore can be omitted.
Also we assume that the relaxation time $\tau$ is independent of 
particle energy. The correction (\ref{eq18})
has the same form as in Eq.~(\ref{eq5}).
Thus, for example,
\begin{eqnarray}
\nonumber
&&v_{f,x} =  <|v|> / \pi - <v_x \eta(v_x)> = \\
\nonumber
&&- \frac{(e m \tau E)^2 }{ 6 (1 + (\omega \tau)^2) } \cos(2 \theta) 
 \int_0^{\infty} v^4  \frac{\partial^2 f_0(v)}{\partial \epsilon^2} dv
\nonumber
\end{eqnarray}
and for the Fermi-Dirac distribution at $T \ll E_F$
we obtain the ratchet flow velocity for the cuts model
\begin{equation}
 \mathbf{v_f} = \frac{\tau^3 e^2 E^2}{ 2 \pi \tau_c \sqrt{2  m^3 E_F} } 
 \frac{1-\pi^2T^2/(24 E_F^2)}{ 1 + (\omega \tau )^2 } 
 \left( \begin{array}{c} 
- \cos(2 \theta) \\ 
2 \sin(2 \theta) 
\end{array} \right)
\label{eq19}
\end{equation}
while for the semidisks model
\begin{eqnarray}
\nonumber
 \mathbf{v_f} = \frac{\pi \tau^3 e^2 E^2}{ 2 (2+\pi) \tau_c \sqrt{2  m^3 E_F} } 
 \frac{1-\pi^2T^2/(24 E_F^2)}{ 1 + (\omega \tau )^2 } 
 \left( \begin{array}{c} 
- \cos(2 \theta) \\ 
  \sin(2 \theta) 
\end{array} \right) .\\
\label{eq20} 
\end{eqnarray}
The angle dependence remains the same as for the Maxwell equilibrium.

The case of weak asymmetry can be also treated in a more 
formal way based of the Green function formalism in the kinetic equation.
First we note that in the second order on alternating electric field the
components of steady-state current density  can be described  by
phenomenological expressions
\begin{eqnarray}
\nonumber
 j_x =\alpha_{xxx}|E_x|^2+ \alpha_{xyy}|E_y|^2,\nonumber\\
 j_y= \mbox{Re}(\alpha_{yxy})(E_xE_y^*+ E_x^*E_y)+
 \mbox{Im}(\alpha_{yxy})[{\bf E} {\bf E}^*]_z \; .
\label{eq21}
\end{eqnarray}
The components of the  photogalvanic tensor $\alpha_{xxx}$, 
$\alpha_{xyy}$ and $\mbox{Re}(\alpha_{yxy})$ determine the
response to the linear-polarized microwave field. For  linear polarization
along $x$ or $y$ axes the current flows along the $x$ direction;
the current in $y$ direction appears for a tilted linear-polarized
electric field.

The kinetic equation written in the operator form reads 
\begin{equation}\label{eq22}
\frac{\partial f}{\partial t} + \hat{F}f= \hat{I}f,
\end{equation}
where  $f(p,\phi)$ is the distribution function $\;\;\;$ and  
${\bf p}=p(\cos{\phi},\sin{\phi})$ is the electron momentum.
 The term ($\hat{F}f$) represents the action of electric field
 ${\bf E}(t) =\mbox{Re}({\bf E}_\omega e^{-i\omega t})$  of the
 electromagnetic wave with the complex amplitude 
${\bf E}_\omega={\bf E}_{-\omega}^*$:
 \begin{eqnarray}\label{eq23}
  && \hat{F}=e{\bf E}(t)\frac{\partial}{\partial {\bf p}}
  =e[E_{x}(\cos{\phi}\frac{\partial}{\partial p}-
\frac{\sin{\phi}}{p}\frac{\partial}{\partial \phi} )+
\nonumber\\&& +E_{y}(\sin{\phi}\frac{\partial}{\partial p}+
\frac{\cos{\phi}}{p}\frac{\partial}{\partial \phi}
)]\equiv\frac{1}{2}\hat{F}_\omega e^{i\omega t}+CC
\end{eqnarray}
The term $\hat{I}f$ is the collision integral
\begin{eqnarray}
\nonumber
 \hat{I}f(p,\phi) = \int_0^{2\pi}
d\phi'[W(\phi',\phi)f(p,\phi')-W(\phi,\phi')f(p,\phi)], \\
\label{eq24}
\end{eqnarray}
where $W(\phi',\phi)$ is the scattering probability.

    The stationary solution of the kinetic equation in the
second order of  electric field  can be written as
\begin{equation}\label{eq25}
    \overline{f}_2=\frac{1}{2}\mbox{Re}
    \Big({\hat{G}^0\hat{F}_\omega\hat{G}^{\omega}\hat{F}_{-\omega} f_0}\Big)
\end{equation}
where $\hat{F}_\omega=e{\bf E}_\omega \frac{\partial}{\partial
{\bf p}}$,
 $\; \hat{G}$ is the Green function of kinetic equation.
For small asymmetry
\begin{equation}\label{eq26}
    \hat{G}^{\omega}=\frac{1}{i\omega-1/\tau}-\frac{1}{(i\omega-1/\tau)^2}\hat{I}^-,
\end{equation}
where $\hat{I}^-$ is the antisymmetric part of scattering operator on
cuts, semidisks etc. The quantity $\tau =\tau_i \tau_c/(\tau_i+\tau_c)$ 
can be attributed to the
symmetric part of scattering on cuts and to the impurity relaxation
time $\tau_i$.

The photogalvanic current reads 
\begin{eqnarray}
\nonumber
&&j_i=\frac{e^3}{4\pi^2}\mbox{Re}\int d{\bf p} \Big\{\tau ^2\hat{I}^-(
    {\bf E}_{-\omega}\frac{\partial}{\partial\bf p})
    \frac{\tau}{1-i\omega \tau}({\bf E}_{\omega}{\bf v})f_0' \\
&&+\tau ({\bf E}_{-\omega}\frac{\partial}{\partial\bf p})\hat{I}^-
    \frac{\tau^2}{(1-i\omega \tau)^2}({\bf E}_{\omega}{\bf v}f_0')\Big\}
\label{eq27}
\end{eqnarray}
Here ${\bf v}={\bf p}/m$ is the electron velocity,
$f_0(\varepsilon)$ is the equilibrium distribution function
in energy $\epsilon$, prime
means derivative over energy.

From Eq.(\ref{eq27}) we find the current induced by 
the linear polarized microwave field
\begin{eqnarray}
\nonumber
&&j_i=\frac{e^3}{4\pi^2}\mbox{Re}\int d{\bf p} \Big\{\tau^2v_i\hat{I}^-v_jv_k
    \Big(\frac{\tau f_0'}{1-i\omega\tau}\Big)' \\
&&- \frac{\tau^2\tau'f_0'}{(1-i\omega\tau)^2}v_iv_j\hat{I}^-v_k\Big\}E_jE_k.
\label{eq28}
\end{eqnarray}

We  use the identities $\hat{I}\xi(\epsilon)=0$,
$<\hat{I}\chi({\bf p})>=0$, and
$<p_ip_j\hat{I}p_k>=-<p_k\hat{I}p_ip_j>$, where $\xi(\epsilon)$ is
an arbitrary function of energy, $\chi({\bf p})$ is an arbitrary
function of momentum; angular brackets stand for the operation of
average over angles in ${\bf p}$-space:
$<...>=\int\frac{d\phi}{2\pi}(...).$ The first and the
second identities follow from the conservation of number of
particles, the third is the consequence of the detailed balance
principle according to which the probability of transition has the 
symmetry $ W({\bf p}',{\bf p})=W(-{\bf p},-{\bf p}')$. 

Further we  consider an algebraic energy dependencies of relaxations
times: $\tau, \tau_c \propto \varepsilon^{s}$. 
This dependence corresponds to the scattering
on a geometric impediment with $\hat{I}^-=(1/\tau_c)\hat{\cal I}^-$
where the operator $\hat{\cal I}^-$ is the integral operator on
$\phi$, $\tau_c$ is the characteristic time of
scattering on cuts or half-disks which depends on the concentration of
scatterers $n_c$.  The power $s=-1/2$ corresponds to scattering
on impurities/scatterers with fixed density in space. We note that
$s=3/2$ may correspond to a case of
charged non-screened impurities distributed in the system plane.
Eq.(\ref{eq28}) can be transformed to the form:
\begin{eqnarray}
\nonumber
&&j_i=\frac{e^3m}{2\pi}\int d\varepsilon f_0'\frac{v^3\tau
    ^3}{\tau_c\varepsilon(1+\omega^2\tau^2)} \Big\{ (2-2s)a_{jki} \\
&&+\frac{s(1-\omega^2\tau^2)}{(1+\omega^2\tau^2)}a_{ijk}\Big\}E_{j}E_{k} \; ,
\label{eq29}
\end{eqnarray}
where ${\bf v}=v{\bf u}$, \ $a_{ijk}=<u_iu_j\hat{\cal I}^-u_k>$
and ${\bf u}$ is a unitary vector.
The stationary current $j_i=\alpha_{ijk}E_{\omega j}E_{\omega
k}^*$ under linear polarized field is given by the following
components of photogalvanic tensor (see Figs.~\ref{fig1},\ref{fig2}): 
$\alpha_{xxx}, \alpha_{xyy}$ and $\mbox{Re}(\alpha_{yxy})$. 
At $T=0$ (a degenerate Fermi gas) the expressions  for these components read
\begin{eqnarray}
\label{eq30}
&&\alpha_{xxx} =
-\frac{e^3V_F\tau^3}{\pi\tau_c(1+\omega^2\tau^2)} \\
\nonumber
&&\times \Bigl[(2-2s)+\frac{s(1-\omega^2\tau^2)}
{(1+\omega^2\tau^2)}\Bigr]a_{xxx} \; ,\\
\label{eq31} 
&&\alpha_{xyy} = \frac{e^3V_F\tau^3}{\pi\tau_c(1+\omega^2\tau^2)} \\
\nonumber
&&\times \Bigl[(2-2s)a_{xxx}-\frac{s(1-\omega^2\tau^2)}
{(1+\omega^2\tau^2)}a_{xyy}\Bigr] \; ,\\
\label{eq32} 
&&\mbox{Re}(\alpha_{yxy})=-\frac{e^3V_F\tau^3}{\pi\tau_c(1+\omega^2\tau^2)}\\
\nonumber
&&\Bigl[(2-2s)a_{xyy}+\frac{s(1-\omega^2\tau^2)}
{2(1+\omega^2\tau^2)}(a_{xyy}-a_{xxx})\Bigr] \; .
\end{eqnarray}
Here, $\tau, \tau_c$ are
relaxation times taken at $\varepsilon=\varepsilon_F$. Now it is
necessary to calculate two quantities $a_{xxx}$ and $a_{xyy}$.
They depend on the model of asymmetric scatterers. For the case of
cuts the scattering probability has the form:
\begin{eqnarray}
\nonumber
&&W(\phi',\phi)={\tau_c}^{-1} [\cos \phi'~\theta(\cos \phi')\delta(\phi'+\phi-\pi) \\
\label{eq33}
&&-{\frac{1}{2}}\cos \phi' \cos \phi~\theta(\cos \phi)\theta(-\cos\phi') ] \; .
\end{eqnarray}
Using Eq.(\ref{eq33}) we obtain    $a_{xxx}=1/48, \; a_{xyy}=-1/16$.
As a result we have in this model:
\begin{eqnarray}
\nonumber
&&\alpha_{xxx} =
-\frac{e^3V_F\tau^3}{48 \pi\tau_c(1+\omega^2\tau^2)^2} \\
\label{eq34}
&&\times [2-s+(2-3s)\omega^2\tau^2] \; ,\\
\nonumber
&&\alpha_{xyy} = \frac{e^3V_F\tau^3}{48 \pi\tau_c(1+\omega^2\tau^2)^2} \\
\label{eq35}
&&\times [2+s+(2-5s)\omega^2\tau^2] \; ,\\
\nonumber
&&\mbox{Re}(\alpha_{yxy})=\frac{e^3V_F\tau^3}{24 \pi\tau_c(1+\omega^2\tau^2)^2} \\
\label{eq36}
&&\times [(3-2s)+(3-4s)\omega^2\tau^2)] \; .
\end{eqnarray}
The formulas (\ref{eq34},\ref{eq35},\ref{eq36}) with
$s=-1/2$ follow also from the exact solution of the problem for cuts
obtained in \cite{entin2006}.

According to the previous Section for 
the semidisks model the scattering probability is
\begin{eqnarray}
\nonumber
&&W(\phi',\phi)=\frac{1}{\tau_c} \biggl\{\cos
\phi'~\theta(\cos \phi')\delta(\phi'+\phi-\pi) \\
\nonumber 
&&+ \frac{1}{4} |\sin (\frac{\phi'-\phi}{2})|\bigl[\theta(\phi-\phi')\theta(-
    \phi'-\phi) \\
\label{eq37}
&&+\theta(\phi'-\phi)\theta(\phi'+\phi)\bigr]\biggr\} \; .
\end{eqnarray}
This probability $W$ leads to the following results for the photogalvanic coefficients
in the semidisks model:
\begin{eqnarray}
\nonumber
&&\alpha_{xxx} = -\alpha_{xyy}= -\mbox{Re}(\alpha_{yxy}) \\
\label{eq38}
&&=-\frac{e^3V_F\tau^3}{12\pi\tau_c(1+\omega^2\tau^2)^2}[2-s+(2-3s)\omega^2\tau^2] \; .
\end{eqnarray}

In the numerical simulations for the cuts model it was taken
that $\tau$ is independent of energy ($s=0$). In such a case
from the above Eqs.~(\ref{eq34},\ref{eq35}\ref{eq36})
and Eq.~(\ref{eq38}) we obtain
\begin{eqnarray}
\nonumber
&&\alpha_{xxx} = -\alpha_{xyy}= -\frac{1}{3}\mbox{Re}(a_{yxy}) \\
\label{eq39}
&&=-\frac{e^3\tau^3\sqrt{2}n_e}{24m\sqrt{m\varepsilon_F}\tau_c(1+\omega^2\tau^2)}
~~~~~ \mbox{(cuts)} ;\\ 
\nonumber
&&\alpha_{xxx} =-\alpha_{xyy}= -\mbox{Re}(a_{yxy}) \\
\label{eq40}
&&=-\frac{e^3\tau^3\sqrt{2}n_e}{6\tau_c
m\sqrt{m\varepsilon_F}(1+\omega^2\tau^2)} ~~~~~~~
(\mbox{semidisks}),
\end{eqnarray}
where $n_e$ is the electron density
related to the current by the relation
${\bf j}=en_e{\bf V_f}$.

Let us also note that at $\tau=const$ for the case of Maxwell
equilibrium we obtain
\begin{eqnarray}
\nonumber
&&\alpha_{xxx} = -\alpha_{xyy}= -\frac{1}{3}\mbox{Re}(a_{yxy}) \\
\label{eq41}
&&=-\frac{\sqrt{2\pi}n_e\tau^3e^3}{24m\sqrt{mT}\tau_c(1+\omega^2\tau^2)}
~~~~~ \mbox{(cuts)}, \\ 
\nonumber
&&\alpha_{xxx} =-\alpha_{xyy}= -\mbox{Re}(a_{yxy}) \\
\label{eq42}
&&=-\frac{\sqrt{2\pi}n_e\tau^3e^3}{6m\sqrt{mT}\tau_c(1+\omega^2\tau^2)}
~~~~~~~ (\mbox{semidisks}) \; .
\end{eqnarray}

\begin{figure}
\vglue 0.5cm
\vglue 0.4cm
   \centering
   \includegraphics[height=0.33\textwidth,angle=0]{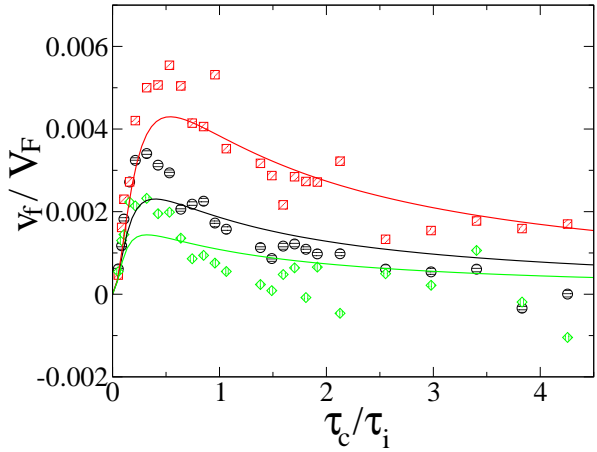}
\caption{Dependence of the rescaled velocity of
ratchet $v_f/V_F$ on the rescaled collision time $\tau_c/\tau_i$
obtained by numerical simulations of the cuts mode
(symbols) for $\omega \tau_i =$ 4.7; 7.05; 9.4
(red, black, green symbols from top to bottom respectively).
The full curves show the theoretical dependence
Eq.~(\ref{eq43},\ref{eq45}) for corresponding $\omega \tau_i$
multiplied by a numerical factor $Q=0.54$.
Here, the equilibrium Fermi-Dirac distribution has
the temperature $T/E_F=0.1$ and $eE \tau_i V_F/E_F=4.2$;
the polarization angle $\theta=0$.
}
\label{fig5}
\end{figure}

It is important to  note that the relation
$\alpha_{xxx}=-\alpha_{xyy}$, which is 
valid for semidisk antidots at any $s$,
implies that the total photogalvanic current
is zero for depolarized radiation.
For a general form of scatterers
and $\tau$ dependent on energy ($s \neq 0$)
even depolarized radiation produce
nonzero photogalvanic current.
This is for example the case for
the cuts model with fixed density
of impurities where $s=-1/2$.

We should again emphasize the difference between formulas obtained
using the kinetic equation approach and simplified way which leads
to Eqs.(\ref{eq6}-\ref{eq7}). The difference in numerical factors
appears due to different methods of averaging:
in Eqs.(\ref{eq6}-\ref{eq7}) it is assumed that the
scattering events are randomly distributed in time,
while in  the kinetic approach it is assumed that
the scattering events are randomly distributed in space.
A simple way to
obtain Eqs.~(\ref{eq39}-\ref{eq42}) is to consider the mean
force applied from the electron gas to an asymmetric scatterer.
This force is determined by the momentum production on the
scatterer, or, in other words, residue of flows of momenta  of
incident $\int v_x {\bf p}Df({\bf p}) d{\bf p}/2\pi^2$
 and scattered particles. If the anisotropy is weak, the
momentum production can be found substituting the expression for
distribution function (\ref{eq18}) into the collision integral
(\ref{eq24}). The flow velocity is determined by equating this force
per unit area to the friction of electron system, $-{\bf P}/\tau$,
where ${\bf P}=m{\bf v}_f n_e$ is the full mean momentum of
electron gas. The result obtained in this way is the same as those 
obtained from the kinetic approach if the mean free time does 
not depend on the energy (Eq. \ref{eq39}-\ref{eq42}).

For a direct comparison with numerical simulations
it is convenient to rewrite Eqs.~(\ref{eq39},\ref{eq40})
to obtain  explicit expressions
for the ratchet velocity. For the Fermi-Dirac distribution
at $T \ll E_F$ this gives 
for the cuts model
\begin{equation}
 \mathbf{v_f} = {\frac{S}{ 24} }  
 \left( \begin{array}{c} 
- \cos(2 \theta) \\ 
3 \sin(2 \theta) 
\end{array} \right)
\label{eq43}
\end{equation}
and for the semidisks model
\begin{equation}
 \mathbf{v_f} = {\frac{S}{ 6} }  
 \left( \begin{array}{c} 
- \cos(2 \theta) \\ 
 \sin(2 \theta) 
\end{array} \right) \; ,
\label{eq44}
\end{equation}
where
\begin{eqnarray}
\nonumber
&&S=\frac{e^2\tau^3\sqrt{2}E^2}{m\sqrt{m E_F}\tau_c(1+\omega^2\tau^2)} \\
\label{eq45}
&&= V_F \frac{(e E \tau V_F)^2 \tau}{2 E_F^2 \tau_c (1+\omega^2 \tau^2)} \; .
\end{eqnarray}

To compare 
the polarization dependence given $\;\;$ 
by Eqs.~(\ref{eq43},\ref{eq44},\ref{eq45})
(see also Eqs.~(\ref{eq41},\ref{eq42})) with the numerical
results shown in Figs.~\ref{fig3},\ref{fig4}
we should take into account that in the numerical simulations
of the cuts model the collisions with cuts take
place always after time $\tau_c$ 
(random positions of cuts in time) while the computations
above assume fixed distribution of scatterers in space
that leads to a factor $3 \sin(2\theta)$ in  Eq.~(\ref{eq43}
instead of  $2 \sin(2\theta)$ in Eqs.~(\ref{eq9},\ref{eq19}).
A part of this, the functional dependence on parameters is
the same in both computations.  A slight difference in a numerical 
coefficient related to different ways of averaging 
for cuts randomly distributed in time (Eq.~(\ref{eq19})) and in space
(Eq.~(\ref{eq43})).

The comparison of the theoretical dependence on 
$\tau, \tau_c$ (Eqs.~(\ref{eq43},\ref{eq45})
with the numerical data in the cuts model is shown in Fig.~\ref{fig5}
at different values of microwave frequency and $\theta=0$.
It shows that the numerical data are well described by 
the theoretical relation
$v_f=-Q S/24$ with a numerical factor $Q \approx 0.5$.
This confirms that the theory
gives a good description of the numerical results.

It is also interesting to compare
the theory for the disks model with
the numerical data of \cite{alik2006} (see Fig.8 there).
Using the value of $\tau_c$ from Eq.~(\ref{eq1})
for a low density at $R/r_d \approx 4$
and $eEr_d/E_F=0.5$ we obtain
from Eqs.~(\ref{eq44},\ref{eq45})
$v_f/V_F \approx 0.14$ instead
of the numerical value
$v_f/V_F \approx 0.05$
(according to Fig.8 in \cite{alik2006}
at $\theta=0$ and $R/r_d=4$ we have
$\omega \tau \approx  0.8$
since $\tau_i \gg \tau_c$).
This means that a numerical factor $Q$
between the theory Eqs.~(\ref{eq44},\ref{eq45})
and the numerical simulations \cite{alik2006} is $Q \approx 1/3$.
Such a correction can be considered as rather
satisfactory since
the theory assumes low density approximation
which is not yet well justified at $R/r_d =4$.
Also the theory of kinetic equation
works in one-collision approximation
which may work not so well for completely
dynamical systems of semidisks lattice.

Finally,
let us return to  a simple derivation of the 
photogalvanic current given in a paragraph after 
Eq.~(\ref{eq42}).
According to this consideration
the directed flow appears on a time scale of the order of $\tau$.
For a purely dynamical system, like
the semidisks Galton board, we have
$\tau_i = \infty$ and $\tau = \tau_c$. We note that $\tau_c$ 
is given by Eq.~(\ref{eq1}) and is proportional to the inverse 
Lyapunov exponent of chaotic dynamics in absence of microwave
driving. This leads to an interesting {\it fundamental}
question about the photogalvanic current
in a purely dynamical system.
Indeed, in absence of friction the semidisks Galton board
with a monochromatic driving is a Hamiltonian system
which has time reversibility property (like $\cos \omega t$),
while the appearance of directed current on a time scale $\tau_c$
breaks time reversibility.
The dependence of the photogalvanic current on 
parameters of a real relaxation process is clear from
the phenomenological point of view:
the current changes its sign with the time inversion
while the electric field squared keeps its sign (Eq.~(\ref{eq21})).
This means that the photogalvanic coefficient $\alpha$ in Eq.~(\ref{eq21})
should contain the relaxation constant. 
By its own, the kinetic equation is irreversible even if
only the static potential scatterers are taken into account.
In fact, a purely dynamical system is reversible and hence
the stationary current should vanish for it.
In principle this point
can be explained by the dynamical
chaos where the time reversibility is broken in practice
for a coarse-grained distribution. However, a more delicate 
point is the question about the detailed balance principle.
In a  dynamical system it means that the transition probabilities
are proportional to a measure in the Hamiltonian phase space
and if all phase space is chaotic there should 
be no global directed current on large time scales. 
In a sense the numerical results \cite{alik2005}
are in favor of this statement since there a small
friction force ${\bf F_f} = -\gamma {\bf p}$
gives the velocity of stationary flow $v_f \sim \gamma R$
which disappears in the limit of $\gamma \rightarrow 0$.
However, in the limit of small $\gamma$ the situation is
somewhat specific since 
the average steady state energy $E_s$, analogous to temperature,
grows with a decrease of $\gamma$ as $E_s \sim ((eE)^2/\gamma)^{2/3} \sim T$
that leads finally to the relations similar to 
those given by Eqs.~(\ref{eq9},\ref{eq15},\ref{eq41},\ref{eq42}).
The numerical simulations performed in \cite{cristadoro}
have been done with the Nos\`e-Hoover dynamics
(see \cite{hoover}) which can be viewed as a purely dynamical 
time reversible Hamiltonian system  
in an extended phase space with additional variables.
Thus the photogalvanic current can appear in a Hamiltonian
asymmetric system with monochromatic driving.
However, the directed flow we discussed takes place 
only in some part of total phase space corresponding to
physical variables and it is possible that
the total current in total extended phase space 
still remains zero. Indeed, in principle it is known that
in chaotic Hamiltonian systems there may be two
separate components (e.g. one with a regular motion
and another with a chaotic motion) with a directed current 
in each component but with the total current 
equal to zero (see examples in \cite{rolandratchet}).
The ratchet analyzed in \cite{cristadoro}
with the Nos\`e-Hoover dynamics can be such a case.
Also, the contradiction can be resolved if to assume 
an existence of a certain time scale
after which  the photogalvanic current is stabilized. 
Thus, from a practical view point we may say that in a dynamical 
chaotic Hamiltonian system signatures of directed
flow appear after the time scale $\tau_c$
but since the current velocity $v_f$ it proportional to a second
power of weak field $v_f \propto (eE)^2$
a relatively long time $t_r \propto 1/(eE)^2$
is needed to observe this current in presence of chaotic fluctuations
and it is necessary that the steady state distribution
in energy is established on a time scale which is shorter than $t_r$.
Further studies are required to understand more deeply the problem
of time reversibility in the context of the photogalvanic effect.

\section{Effects of a magnetic field for the semidisks Galton board}

For experiments on photogalvanic current in asymmetric nanostructures
it is important to know what are the effects of a magnetic
field $B$ perpendicular to the 2DES plane on the strength
of current and its directionality.
An analytic solution of the kinetic equation
becomes much more complicated compared to the cases
considered above. This is especially the case when the
Larmor radius $R_L$ of electron motion becomes comparable
with the size of asymmetric antidots.  
Therefore, the numerical simulations in this case
become especially important.
For the semidisks Galton board the effects of magnetic field
have been studied in \cite{alik2006}.
They clearly show that the ratchet current becomes quite
weak when the Larmor radius $R_L$ becomes smaller 
then the semidisk radius $r_d$.
This is rather natural from a physical view point
since in this regime the scattering on semidiscs is suppressed.

\begin{figure}
\vglue 0.5cm
\vglue 0.4cm
   \centering
   \includegraphics[height=0.33\textwidth,angle=0]{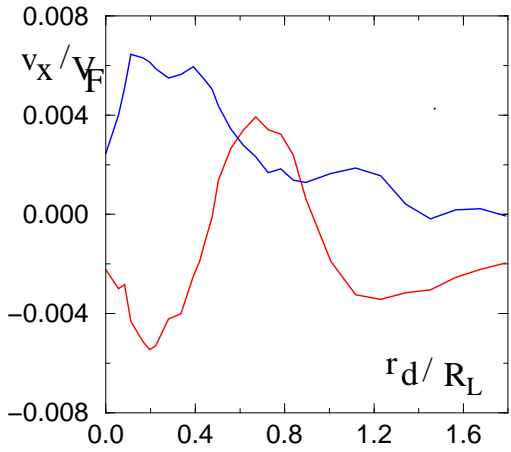}
\caption{Dependence of rescaled ratchet velocity $v_x/V_F$ 
in $x-$direction
on the ratio $r_d/R_L $ 
proportional to the magnetic field $B$ perpendicular
to the semidisks Galton board. The system parameters are
$R/r_d=2.5$, $T/E_F=0.1$, $eEr_d/E_F=0.3$,
$l_\omega/r_d=2\pi V_F/(\omega r_d) = 9$.
The bottom (top) red (blue) curve corresponds to
the polarization angle $\theta=0$ ($\theta=\pi/2$).
The rescaled mean free path due to impurity scattering
is $l/r_d=V_F \tau_i/r_d=45$; the rescaled computation time is
$\omega t = 3.1 \times 10^6$.
}
\label{fig6}
\end{figure}

\begin{figure}
\vglue 0.5cm
\vglue 0.4cm
   \centering
   \includegraphics[height=0.33\textwidth,angle=0]{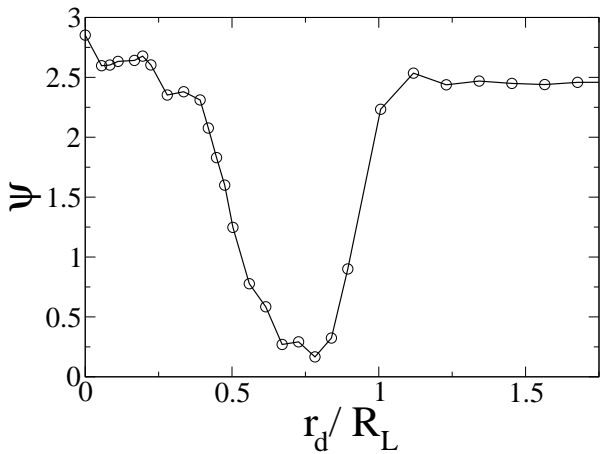}
\caption{Dependence of ratchet current angle $\psi$
on a rescaled magnetic field $r_d/R_L \propto B$
for the case of Fig.~\ref{fig6} at $\theta=0$;
curve with circles shows numerical data,
angle is given in radians.
}
\label{fig7}
\end{figure}

However, in the regime with $r_d/R_L \sim 1$
a relatively weak magnetic field can significantly
affect the directionality of photogalvanic current.
This is illustrated in Figs.~\ref{fig6},\ref{fig7}
obtained by numerical simulations with the method
described in \cite{alik2006}. The results
of these Figures clearly show that a
moderate magnetic field can change the
direction of current almost on 180 degrees
(Fig.~\ref{fig7}). The angular dependence of Fig.~\ref{fig7}
is not sensitive to the microwave field strength
and therefore is not related to the Lorentz force.
We attribute the origin of this strong angular dependence 
to a significant change of scattering process in the
regime when   $r_d/R_L \sim 1$ related to multiple
collisions of electron with a semidisk.

\begin{figure}
\vglue 0.5cm
\vglue 0.4cm
   \centering
   \includegraphics[height=0.33\textwidth,angle=0]{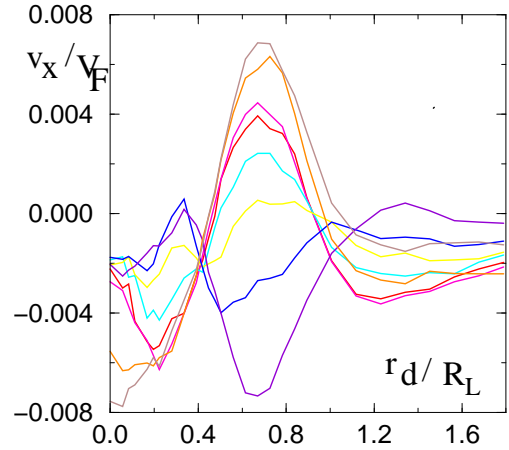}
\caption{Dependence of the ratchet rescaled velocity
$v_x/V_F$ on the rescaled ratio $r_d/R_L \propto B$
for parameters of Fig.~\ref{fig6} at $\theta=0$
and various values of rescaled microwave frequency $\omega r_d/V_F$=
0.335 (brown), 0.447 (orange), 0.648 (magenta),
0.693 (red), 0.805 (cyzen), 0.939 (yellow),
1.12 (blue), 1.39 (violet)
(curves from top to bottom at $r_d/R_L=0.8$).
}
\label{fig8}
\end{figure}

To make a more close link to possible 
experimental studies we note that
for the electron density $n_e = 2.5 \times 10^{11} cm^{-2}$,
an effective electron mass $m=0.067m_e$
and the semidisk radius $r_d=0.4 \mu m$
we have for the parameters of Fig.~\ref{fig6}
the following physical values:
$E_F \approx 100 K$, 
$V_F = 2.2 \times 10^7 cm/s$, 
$\omega/2\pi = 60GHz$, $l_\omega = 2\pi V_F/\omega = 3.6 \mu m$ 
and $l = V_F \tau_i = 18 \mu m$
(such value of $l$ corresponds to
mobility of about $2 \times 10^6 cm^2/V S$), a magnetic field
$B=0.075 T$ corresponds to 
the Larmor frequency $\omega_L \approx 34 GHz$ and
$r_d/R_L =0.4$. For these parameters and data of Fig.~\ref{fig8}
the microwave frequency changes from
$29GHz$ to $120GHz$ when the rescaled ratio changes
from $\omega r_d/V_F=0.335$ to  1.39
($\omega r_d/V_F=0.693$ for Fig.~\ref{fig6}). 

In absence of magnetic field there is no dependence of current direction
on the microwave frequency. On the contrary, in the regime $r_d/R_L \sim 1$
the directionality of flow can be also changed by changing $\omega$
as it is shown in Fig.~\ref{fig8}. We also note that the velocity
of flow in $x$ direction remains the same 
with a change $B \rightarrow -B$ due to symmetry reasons
and we present data only for $B>0$.

\begin{figure}
\vglue 0.5cm
\vglue 0.4cm
   \centering
   \includegraphics[height=0.33\textwidth,angle=0]{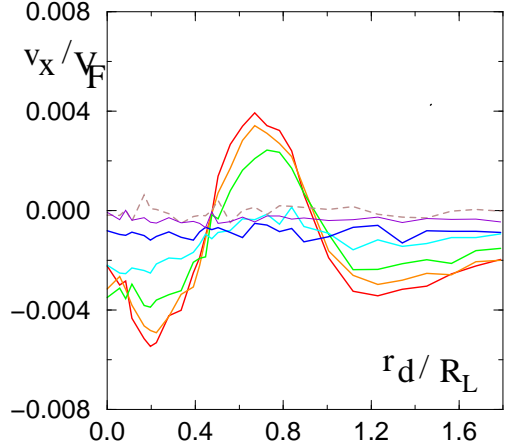}
\caption{Dependence of the ratchet rescaled velocity
$v_x/V_F$ on the rescaled ratio $r_d/R_L \propto B$
for parameters of Fig.~\ref{fig6} at $\theta=0$
and various values of rescaled mean free path
$l/r_d=V_F \tau_i/r_d=$ 45, 22.5, 11.1, 4.5, 2.25, 1.1
(full curves from top to bottom at $r_d/R_L=0.8$;
for the physical parameters given in the text $l$
changes from 18$\mu m$ to 0.45$\mu m$);
dashed curve corresponds to the case when semidisks
are replaced by disks and $l/r_d=45$.
}
\label{fig9}
\end{figure}

This angular dependence becomes weaker when the mean free
path $l$ decreases due to decrease of impurity scattering time $\tau_i$
(see Fig.~\ref{fig9}). The results of Fig.~\ref{fig9}
also show that the ratchet effect disappears
when $l$ becomes smaller than size of asymmetric antidot.
Of course, the current is absent when
the semidisks are replaced by disks
(in this case the numerical data,
shown by dashed curve in Fig.~\ref{fig9}, are on 
a level of statistical fluctuations). 

The angular dependence of the photogalvanic current
$B$ and $\omega$ found in this Section is rather
nontrivial and 
further studies are required 
to obtain a detailed explanation for it.

\section{Discussion}
In this paper we developed a theory which
determines the strength
and directionality of the phogalvanic current in
artificial asymmetric nanostructures. The theoretical
results are in good agreement with detailed
numerical simulations performed in this work and
in \cite{alik2006}. We also find from numerical simulations that
the directionality of photogalvanic current
is very sensitive to weak magnetic fields.

A microwave field of $E = 1 V/cm$ generates 
in a lattice of semidisks with  $R \sim 1 \mu m$
a current of about  0.2 nA per lattice row
at electron density $n_e \approx 2.5 \times 10^{11}$.
For a structure  of 100$\mu m$ we have currents
of about 20 nA. The photogalvanic effect
has classical grounds and exists at
room temperature if the mean free path
remains larger (or comparable) than the size
of asymmetric antidot. This is the case
of the Lund experiment \cite{song} with 
$l \sim R \sim 100 n m$ 
(the polarization there corresponds to $\theta=\pi/2$,
triangles are used instead of disks).
For a fixed ratio $R/r_d$ 
a decrease of $R \sim l$ by a factor 10 gives a drop
of current by a factor 100, however, 
the induced voltage drops only by a factor 10
since a resistance also drops 10 times.
While the experiments \cite{lorke,song}
demonstrate the existence of the effect
there are still no experimental
data to be compared with
our theoretical and numerical results:
in the experiments the strength of current
is known but  a microwave field strength
acting on electrons is not well defined.
Also till now only two polarization cases
have been analyzed experimentally
($\theta=0$ in \cite{lorke} and
$\theta=\pi/2$ in \cite{song}).
The directionality of current in these
experiments is in agreement with our
results but a detailed 
experimental investigation of the polarization dependence is
still highly desirable.

All these considerations  clearly show that
further experimental studies of photogalvanic
currents in asymmetric nanostructures are very interesting
and important. Indeed, the theory developed above 
considers only the case of noninteracting particles.
In reality electron-electron interaction
may play an important role, as well as 
plasmon modes can also affect the photogalvanic effect.
Also at small scales with $R \sim 100 n m$
the level spacing between quantum levels
inside one lattice cell becomes of the order of
$100 GHz$ (at $n_e \sim 3 \times 10^{11}$)
and quantum effects may play an important role.
Also at $R < 100 n m$ the collision frequency enters in
teraherz range $V_F/R > 1 THz$. Thus such asymmetric nanostructures
can be used as room temperature detectors
of radiation in teraherz range. In analogy with the
GORE-TEX material used in common life
for one-way transport of water and air \cite{goretex}
such artificial asymmetric nanostructures
may be considered as ``NANO-GORE-TEX'' material.
The further studies of the NANO-GORE-TEX properties
may bring new interesting applications
including high frequency detectors and sensors.

We thank Z.D.Kvon  for useful discussions.
This work was supported in part by the ANR PNANO project MICONANO
and (for MVE and LIM) by the grant of RFBR No 04-02-16398, Program
for support of scientific schools of the
Russian Federation No. 4500.2006 and INTAS No. 03-51-6453.

\end{document}